\documentstyle[12pt]{article}
\begin{document}
\begin{titlepage}
\begin{flushright}
IC/2001/69\\
hep-th/0107018
\end{flushright}
\vspace{10 mm}

\begin{center}
{\Large Bulk Fields in Delocalized Dilatonic $p$-Branes}

\vspace{5mm}

\end{center}
\vspace{5 mm}

\begin{center}
{\large Donam Youm\footnote{E-mail: youmd@ictp.trieste.it}}

\vspace{3mm}

ICTP, Strada Costiera 11, 34014 Trieste, Italy

\end{center}

\vspace{1cm}

\begin{center}
{\large Abstract}
\end{center}

\noindent

We study localization properties of various bulk fields on a dilatonic 
$p$-brane which is delocalized along its transverse directions except one.  
We find that all the bosonic and fermionic bulk fields can be localized on 
the delocalized dilatonic $p$-brane in a strict sense, namely the 
Kaluza-Klein zero modes of the bulk fields are normalizable and are localized 
around the brane, for any values of the dilaton coupling parameter.  

\vspace{1cm}
\begin{flushleft}
July, 2001
\end{flushleft}
\end{titlepage}
\newpage
\begin{sloppypar}

Recently, there has been active interest in brane world scenarios 
\cite{add,aadd,rs1,rs2}, because they provide possible attractive solutions 
to the hierarchy and the cosmological constant problems.  In brane world 
scenarios, it is generally assumed that only gravity lives in the bulk and 
all the fields of the Standard Model are confined within the brane 
worldvolume.  Such point of view may be natural in string theories, where 
matter fields on the brane can be identified as open string modes and the 
gravity in the bulk is a closed string mode.  But when we consider the 
embeddings of brane world scenarios into supergravity theories, we usually 
encounter various bulk matter fields, especially those compactified from 
10 or 11 dimensions if we are interested in effective supergravity 
descriptions of string theories.  So, it would be of interest to study 
various bulk matter fields in brane world models.  (The previous studies 
on various bulk fields in the brane world scenarios can be found, for 
example, in Refs. \cite{gw,dhr,pom,bg,grn,chn,iod,rds,kho,tac}.)  

In the case of the Randall-Sundrum (RS) domain wall with noncompact extra 
space \cite{rs2}, it was shown \cite{dhr,pom,bg,grn,chn,iod,rds} that it is 
impossible to localize bulk gauge fields and bulk fermionic fields on the 
domain wall through purely gravitational mechanism.  (Cf. See also Refs. 
\cite{kss,sil,dl}.)  It was also proposed that a $U(1)$ field on the brane 
originates rather from two bulk two-form potentials \cite{pl,dls} and 
alternative methods for localizing the bulk $U(1)$ field were also proposed 
\cite{oda,gn}.  In our previous works \cite{youm1,youm2,youm3}, we showed 
that dilatonic domain walls can also localize bulk gravity, provided the 
tension of the wall is positive.  Unlike the (nondilatonic) RS domain wall, 
the dilatonic domain walls are shown \cite{youm2} to localize all the bosonic 
and fermionic bulk fields, provided the dilaton coupling parameter is large 
enough in some cases.  However, some of bulks fields are not localized on 
the dilatonic domain walls in a strict sense.  Namely, although the 
Kaluza-Klein (KK) zero modes of all the bulk fields are normalizable, the 
KK zero modes of some bulk fields are evenly distributed along the extra 
spatial direction or localized away from the wall.  Another problem with 
the dilatonic domain walls which localize gravity is that there are naked 
singularities at finite distance away from the wall.  

In Ref. \cite{youm4}, we studied localization properties of bulk gravity on 
dilatonic $p$-branes in arbitrary spacetime dimensions.  We showed that 
generally bulk gravity cannot be localized on a $p$-brane (with the 
codimension higher than or equal to three) unless the transverse space is 
truncated, for example, through the introduction of another $p$-brane at 
finite distance from the existing one.  However, when all the transverse 
directions except one are delocalized, the bulk gravity is shown to be 
localized on the brane, if a parameter in the $p$-brane harmonic function 
is chosen such that there are curvature singularities on both sides of the 
brane at finite coordinate distance.  Such singularities of the delocalized 
$p$-branes are less harmful than the naked singularities of the dilatonic 
domain wall in a sense that massive matter with finite energy can never 
reach the singularities, being repelled by the singularities.  The warp 
factor, which sets the energy scale on the hypersurface, of such delocalized 
brane solution increases as one moves away from the brane, asymptotically 
approaching infinity as the singularities are approached.  So, one may 
regard such delocalized brane as the TeV brane.  In this note, we study 
the localization properties of various bulk fields on the delocalized 
dilatonic $p$-branes.  We find that all the bosonic and fermionic bulk 
fields are localized on the brane in a strict sense, namely not only the 
KK zero mode is normalizable but also its field distribution is localized 
around the brane.  

We begin by discussing delocalized dilatonic $p$-brane solution studied in 
Ref. \cite{youm4}.  Just as in Ref. \cite{youm4}, we regard the $p$-brane 
as a solitonic brane magnetically charged under the field strength $F_n$ of 
the rank $n=D-p-2$, although such assumption is not essential for our 
discussion.  The action for a dilatonic $p$-brane in $D$-dimensional 
spacetime with an arbitrary dilaton coupling parameter $a_p$ is
\begin{equation}
S_p={1\over{2\kappa^2_D}}\int d^Dx\sqrt{-G}\left[{\cal R}-{4\over{D-2}}
(\partial\phi)^2-{1\over{2\cdot n!}}e^{-2a_p\phi}F^2_n\right].
\label{action}
\end{equation}
The solution for the extreme dilatonic $p$-brane with the longitudinal 
coordinates ${\bf x}=(x^1,\dots,x^p)$ and the transverse coordinates ${\bf y}
=(y^1,\dots,y^{n+1})$, located at ${\bf y}={\bf 0}$, has the following form:
\begin{eqnarray}
G_{MN}dx^Mdx^N&=&H^{-{{4(n-1)}\over{(p+n)\Delta_p}}}_p\left[-dt^2+d{\bf x}
\cdot d{\bf x}\right]+H^{{4(p+1)}\over{(p+n)\Delta_p}}_pd{\bf y}\cdot d{\bf y},
\cr
e^{2\phi}&=&H^{{(p+n)a_p}\over{\Delta_p}}_p,\ \ \ \ \ \ \ 
F_n=\star(dH_p\wedge dt\wedge dx_1\wedge\dots\wedge dx_p),
\cr
H_p&=&1+{Q_p\over{|{\bf y}|^{n-1}}},\ \ \ \ \ \ \ \ 
\Delta_p\equiv{{(p+n)a^2_p}\over 2}+{{2(p+1)(n-1)}\over{p+n}}.
\label{pbrnsol}
\end{eqnarray}
Such dilatonic $p$-brane covers all the possible single charged $p$-branes 
in string theories.  We assume that the $p$-brane is delocalized in the 
$n$-dimensional subspace ${\cal K}_n$ with the coordinates $y^i$ 
($i=1,\dots,n$) of the transverse space, where ${\cal K}_n$ is a Ricci flat 
compact manifold with the metric $d\tilde{s}^2_n=g_{ij}(y^k)dy^idy^j$.  Then, 
the metric of the delocalized solution has the following form:
\begin{equation}
G_{MN}dx^Mdx^N=H^{-{{4(n-1)}\over{(p+n)\Delta_p}}}_p\left[-dt^2+d{\bf x}
\cdot d{\bf x}\right]+H^{{4(p+1)}\over{(p+n)\Delta_p}}_p\left[d\tilde{y}^2
+g_{ij}dy^idy^j\right],
\label{dlcmet}
\end{equation} 
where $\tilde{y}:=y^{n+1}$ and the harmonic function is now given by $H_p=
1+\tilde{Q}_p|\tilde{y}|$.  When $\tilde{Q}_p<0$, the zeros $\tilde{y}=
\pm\tilde{Q}^{-1}_p$ of $H_p$ correspond to the curvature singularities.  
When $\tilde{Q}_p>0$, there is no singularity except at $\tilde{y}=0$, 
where the delocalized $p$-brane is located.  For the purpose of studying 
the KK modes of bulk fields, it is convenient to transform the transverse 
coordinate $\tilde{y}$ to a new one $y$ with which the $(p+2)$-dimensional 
part of the metric (with the coordinates ($x^{\mu},y$)) takes the standard 
form of the RS model with the warp factor ${\cal W}$.  We showed \cite{youm4} 
that bulk graviton can be localized on the delocalized $p$-brane when there 
are singularities on both sides of the brane.  For such case, the metric in 
the new coordinates is given by
\begin{eqnarray}
G_{MN}dx^Mdx^N&=&{\cal W}\left[-dt^2+d{\bf x}\cdot d{\bf x}\right]
+dy^2+{\cal W}^{-{{p+1}\over{n-1}}}g_{ij}dy^idy^j,
\cr
{\cal W}&=&\left[1-K_p|y|\right]^{-{{4(n-1)}\over{2(p+1)+(p+n)\Delta_p}}}
=\left[1-K_p|y|\right]^{-{{8(n-1)}\over{(p+n)^2a^2_p+4(p+1)n}}},
\cr
K_p&\equiv&-{{2(p+1)+(p+n)\Delta_p}\over{(p+n)\Delta_p}}\tilde{Q}_p
\cr
&=&-{{(p+n)^2a^2_p+4(p+1)n}\over{(p+n)^2a^2_p+4(p+1)(n-1)}}\tilde{Q}_p.
\label{warpmet}
\end{eqnarray}
Note, the parameter $K_p$ is always positive when $\tilde{Q}_p<0$.  

First, we study the localization properties of bosonic bulk fields.  All 
the bosonic fields can be collectively considered as $q$-form potentials, 
since scalar and $U(1)$ gauge fields can be respectively regarded as 0- and 
1-form potentials.  The action for a massless $q$-form potential 
$A_{M_1\dots M_q}$ with the field strength $F_{M_1\dots M_{q+1}}=(q+1)
\partial_{[M_1}A_{M_2\dots M_{q+1}]}$ in the $D$-dimensional bulk spacetime 
is given by
\begin{equation}
S_q=-{1\over{2\cdot(q+1)!}}\int d^Dx\sqrt{-G}G^{M_1N_1}\cdots G^{M_{q+1}
N_{q+1}}F_{M_1\dots M_{q+1}}F_{N_1\dots N_{q+1}},
\label{qformact}
\end{equation}
from which we obtain the following equations of motion for $A_{M_1\dots M_q}$:
\begin{equation}
{1\over\sqrt{-G}}\partial_{M_1}\left[\sqrt{-G}G^{M_1N_1}\cdots G^{M_{q+1}
N_{q+1}}F_{N_1\dots N_{q+1}}\right]=0.
\label{qfrmeq}
\end{equation}
We take the following KK mode ansatz for the bulk $q$-form potential:
\begin{equation}
A_{\mu_1\dots\mu_q}(x^{\mu},y,y^i)=a^{(m)}_{\mu_1\dots\mu_q}(x^{\mu})u_m(y),
\label{kkqfrm1}
\end{equation}
where $a^{(m)}_{\mu_1\dots\mu_q}$ is assumed to satisfy the following field 
equations for a massive $q$-form potential in $(p+1)$-dimensional flat 
spacetime:
\begin{equation}
\partial^{\mu_1}f^{(m)}_{\mu_1\dots\mu_{q+1}}+m^2a^{(m)}_{\mu_2\dots
\mu_{q+1}}=0,
\label{fltqfeq}
\end{equation}
along with the gauge condition $\partial^{\mu_1}a^{(m)}_{\mu_1\dots\mu_q}=0$, 
where $f^{(m)}_{\mu_1\dots\mu_{q+1}}\equiv(q+1)\partial_{[\mu_1}a^{(m)}_{\mu_2
\dots\mu_{q+1}]}$ is the field strength of $a^{(m)}_{\mu_1\dots\mu_q}$.  
Then, from Eq. (\ref{qfrmeq}) we obtain the following Sturm-Liouville 
equation satisfied by $u_m(y)$:
\begin{equation}
\partial_y\left[{\cal W}^{-{{p+1}\over{2(n-1)}}-q}\partial_yu_m\right]
=m^2{\cal W}^{-{{p+1}\over{2(n-1)}}-q-1}u_m,
\label{qfmsl1}
\end{equation}
from which we see that the eigenfunctions $u_m$ with different eigenvalues 
$m^2$ are orthogonal to one another w.r.t. the weighting function $w(y)=
{\cal W}^{-{{p+1}\over{2(n-1)}}-q-1}$, provided appropriate boundary 
conditions on $u_m$ and $u^{\prime}_m$ are satisfied.  In the case of the 
KK zero mode ($m=0$), the general solution to Eq. (\ref{qfmsl1}) is given by
\begin{equation}
u_0(y)=c_1+c_2{\cal W}^{{{p+1}\over{2(n-1)}}+q-{{(p+n)^2a^2_p+4(p+1)n}
\over{8(n-1)}}},
\label{kkzero1}
\end{equation}
where $c_1$ and $c_2$ are integration constants.  It can be shown that 
the $y$-dependent term in Eq. (\ref{kkzero1}) gives rise to the mass term 
in the effective action of the $q$-form potential, which is contradictory 
to the fact that $u_0$ is the KK zero mode.  Furthermore, the boundary 
condition on $\tilde{u}^{\prime}_0$, where $\tilde{u}_0\equiv{\cal 
W}^{-{{p+1}\over{4(n-1)}}-{q\over 2}}u_0$, at $y=0$, resulting from the 
$\delta$-function potential in the Schr\"odinger equation satisfied by 
$\tilde{u}_0$, requires that $c_2=0$.  So, the consistent KK zero mode is 
independent of $y$.  This KK zero mode is normalizable, if its norm
\begin{equation}
\int^{1/K_p}_{-1/K_p}dy\,u^2_0(y)w(y)\sim\left.
(1-K_p|y|)^{{8(q+1)(n-1)+4(p+1)(n+1)+(p+n)^2a^2_p}\over{(p+n)^2a^2_p
+4(p+1)n}}\right|^{1/K_p}_{-1/K_p}
\label{normcnd1}
\end{equation}
is finite, which is the case when 
\begin{equation}
8(q+1)(n-1)+4(p+1)(n+1)+(p+n)^2a^2_p>0.
\label{qfrmrmcd}
\end{equation}  
Note, the $n=0$ case corresponds to the dilatonic domain wall (with 
codimension one).  For such case, for the KK zero mode of the form potential 
with the rank $q>(p-1)/2$ to be normalizable, a large enough value of $a_p$ 
is required \cite{youm5}, as can be seen also from Eq. (\ref{qfrmrmcd}).  
However, when the codimension of the brane is higher than one ($n\geq 1$), 
the KK zero mode of bulk form potential of any ranks $q\leq p$ is 
normalizable for any value of the dilaton coupling parameter $a_p$.  
We proposed in Ref. \cite{youm5} that we have to consider the quantity 
$u^2_0(y)w(y)$ for determining distribution of the bulk field KK zero mode 
across the extra spatial direction.  With this criterion we see that field 
distribution of the KK zero mode of the bulk form potential of any ranks is 
localized around the brane (with the codimension higher than one) for any 
values of $a_p$.  So, the delocalized $p$-branes localize bulk form 
potentials of any ranks in a strict sense, namely, the KK zero mode is 
normalizable and its field distribution is localized around the brane.  

For completeness, we study the dimensional reduction of the bulk form 
potential to a form potential of lower rank.  First, we consider the KK zero 
mode ansatz of the following form:
\begin{equation}
A_{\mu_1\dots\mu_ri_1\dots i_{q-r}}(x^{\mu},y,y^i)=a_{\mu_1\dots\mu_r}
(x^{\mu})u(y),
\label{qfrmkk2}
\end{equation}
with all the other components vanishing, where $q-r\leq n$.  Following the 
similar procedure as above, we find that the KK modes are orthogonal w.r.t. 
the weighting function $w(y)={\cal W}^{{{(2q-2r-1)(p+1)}\over{2(n-1)}}-r-1}$ 
and the consistent KK zero mode is independent of $y$.  So, the KK zero 
mode is normalizable, if the following condition is satisfied:
\begin{equation}
8(r+1)(n-1)+4(p+1)(n+1-2q+2r)+(p+n)^2a^2_p>0.
\label{qfmnmcd2}
\end{equation}
As long as $q-r<(r+1)(n-1)/(p+1)+(n+1)/2$, the KK zero mode is normalizable 
for any values of $a_p$.  Otherwise, large enough value of $a^2_p$ is 
required for normalization of the KK zero mode.  By considering the 
quantity $u^2_0(y)w(y)$, we see that the field distribution for the KK 
zero mode is localized around the brane, when
\begin{equation}
2(r+1)(n-1)+(p+1)(1-2q+2r)>0.
\label{pfrmkkzlc1}
\end{equation}
For such case, the KK zero mode is also normalizable for any values of 
$a_p$ and therefore the brane localizes the bulk $q$-form potential in 
a strict sense.  Second, we consider the following form of the KK zero 
mode ansatz:
\begin{equation}
A_{\mu_1\dots\mu_ryi_1\dots i_{q-r-1}}(x^{\mu},y,y^i)=a_{\mu_1\dots\mu_r}
(x^{\mu})u(y),
\label{qfrmkk3}
\end{equation}
with all the other components taken to vanish, where $q-r-1\leq n$.  By 
substituting this ansatz into Eq. (\ref{qfrmeq}), we obtain the following 
equation satisfied by $u(y)$:
\begin{equation}
\partial_y\left[{\cal W}^{{{(p+1)(2q-2r-3)}\over{2(n-1)}}-r-1}u\right]=0,
\label{eqfru}
\end{equation}
from which we obtain $u\sim{\cal W}^{{{(p+1)(3-2q+2r)}\over{2(n-1)}}+r+1}$.  
To determine the normalizability, we substitute this KK zero mode ansatz 
into the bulk action (\ref{qformact}).  The resulting action has the 
following form:
\begin{eqnarray}
S_q&\sim&-{1\over{2\cdot(r+1)!}}\int^{1/K_p}_{-1/K_p}dy
(1-K_p|y|)^{{4(2q-2r-3)(p+1)-8(r+1)(n-1)}\over{(p+n)^2a^2_p+4(p+1)n}}
\cr
& &\times\int d^{p+1}x\,\eta^{\alpha_1\beta_1}\cdots\eta^{\alpha_{r+1}
\beta_{r+1}}f_{\alpha_1\dots\alpha_{r+1}}f_{\beta_1\dots\beta_{r+1}},
\label{rfrmeffact}
\end{eqnarray}
from which we see that the KK zero mode is normalizable when
\begin{equation}
4(n-3+2q-2r)(p+1)-8(r+1)(n-1)+(p+n)^2a^2_p>0.
\label{pfrmnmalcd3}
\end{equation}
So, when $q-r-1>(n-1)(2r-p+1)/2(p+1)$, the KK zero mode is normalizable 
for any $a_p$.  Otherwise, large enough $a^2_p$ is required for the 
normalization.  From the integrand of the $y$ integral in Eq. 
(\ref{rfrmeffact}), we see that furthermore the field distribution of 
the KK zero mode is localized around the brane when $q-r-3/2>(r+1)
(n-1)/(p+1)$, for which case the brane localizes the bulk $q$-form potential 
in a strict sense.  

Next, we study the localization properties of bulk fermionic fields.  
First, we consider the massless bulk spin-1/2 fermion $\Psi$ with the action
\begin{equation}
S_{1/2}=\int d^Dx\sqrt{-G}\,i\bar{\Psi}\Gamma^MD_M\Psi,
\label{spnact}
\end{equation}
where $D_M\equiv\partial_M+{1\over 4}w_{MAB}\gamma^{AB}$ is the gravitational 
covariant derivative on a spinor.  Here, $\omega_{MAB}$ is the spacetime 
spin-connection associated with the bulk metric $G_{MN}$ and $\gamma^{AB}
\equiv\gamma^{[A}\gamma^{B]}$.  The convention for the spacetime vector 
indices are $M,N,P,...$ [$A,B,C,...$] for the curved [flat-tangent] bulk 
spacetime indices, $\mu,\nu,\rho,...$ [$\alpha,\beta,\gamma,...$] for the 
curved [flat-tangent] worldvolume spacetime indices, and $i,j,k,...$ 
[$a,b,c,...$] for the curved [flat-tangent] transverse space indices.  The 
gamma matrices $\Gamma^M\equiv E^M_A\gamma^A$ with curved indices satisfy 
$\{\Gamma^M,\Gamma^N\}=2G^{MN}$, where $E^M_A$ is the inverse of the vielbein 
$E^A_M$.  From the bulk action (\ref{spnact}),  we obtain the following 
equation of motion for the bulk spinor field:
\begin{equation}
\Gamma^M(\partial_M+{1\over 4}\omega_{MAB}\gamma^{AB})\Psi=0,
\label{spneq}
\end{equation}
which reduces to the following form after the bulk metric (\ref{dlcmet}) is 
substituted:
\begin{equation}
\left[\Gamma^y\left(\partial_y-{{p+1}\over{4(n-1)}}{{{\cal W}^{\prime}}\over
{\cal W}}\right)+\Gamma^{\mu}\partial_{\mu}+\Gamma^m\tilde{D}_m\right]\Psi=0,
\label{spneq2}
\end{equation}
where $\tilde{D}_m$ is the gravitational covariant derivative associated 
with the metric $g_{mn}(y^i)$.  In general, the KK mode ansatz for the 
bulk spinor has the form
\begin{equation}
\Psi(x^{\mu},y,y^i)=\psi(x^{\mu})u(y)\chi(y^i).
\label{spnkk}
\end{equation}
Since we are interested in finding the KK zero mode, we impose the zero 
mode conditions $\gamma^{\alpha}\partial_{\alpha}\psi=0$ and $\Gamma^m
\tilde{D}_m\chi=0$.  Then, the bulk equation of motion (\ref{spneq2}) 
reduces to the following equation satisfied by $u(y)$:
\begin{equation}
\Gamma^y\left[\partial_y-{{p+1}\over{4(n-1)}}{{{\cal W}^{\prime}}\over
{\cal W}}\right]u=0,
\label{frmzrmeq}
\end{equation}
from which we have $u\sim{\cal W}^{{p+1}\over{4(n-1)}}$.  
To check the normalizability, we plug this KK zero mode into the bulk action:
\begin{eqnarray}
\int d^Dx\sqrt{-G}\,i\bar{\Psi}\Gamma^MD_M\Psi\sim
\int^{1/K_p}_{-1/K_p}dy{\cal W}^{-{1\over 2}}\int d^{p+1}x\,i\bar{\psi}
\gamma^{\alpha}\partial_{\alpha}\psi
\cr
=\int^{1/K_p}_{-1/K_p}dy\left(1-K_p|y|\right)^{{4(n-1)}\over{(p+n)^2a^2_p
+4(p+1)n}}\int d^{p+1}x\,i\bar{\psi}\gamma^{\alpha}\partial_{\alpha}\psi.
\label{frmkkzrrm}
\end{eqnarray}
Note, in the case of the dilatonic domain walls \cite{youm2}, a large enough 
value of the dilaton coupling parameter is required for localizing the bulk 
spinor field, as can be seen also from this equation with $n=0$.  On the 
other hand, in the case of the delocalized $p$-brane (with $n\geq 1$), the 
KK zero mode of the bulk fermion is normalizable for any $a_p$.  
Furthermore, when $n>1$, the field distribution of the KK zero mode is 
localized around the brane as well, so the brane localizes the bulk fermion 
in a strict sense.  

Second, we consider the bulk gravitino $\Psi_M$ with the action
\begin{equation}
S_{3/2}=\int d^Dx\sqrt{-G}\,i\bar{\Psi}_M\Gamma^{MNP}D_N\Psi_P,
\label{grtnact}
\end{equation}
where $D_M\Psi_N\equiv\partial_M\Psi_N-\Gamma^P_{MN}\Psi_P+{1\over 4}
\omega_{MAB}\gamma^{AB}\Psi_N$.  Here, $\Gamma^P_{MN}$ is the affine 
connection associated with $G_{MN}$ and $\Gamma^{MNP}\equiv\Gamma^{[M}
\Gamma^N\Gamma^{P]}$.  So, the equation of motion for $\Psi_M$ is
\begin{equation}
\Gamma^{MNP}D_N\Psi_P=0.
\label{grtneq}
\end{equation}
We choose the gauge $\Psi_y=0$ and assume $\Psi_m=0$.  In general, the 
KK mode ansatz then has the form:
\begin{equation}
\Psi_{\mu}(x^{\nu},y,y^i)=\psi_{\mu}(x^{\nu})u(y)\chi(y^i).
\label{grtnkk}
\end{equation}
To obtain the KK zero mode of the bulk gravitino, we impose the zero mode 
conditions $\gamma^{\alpha\beta\delta}\partial_{\beta}\psi_{\delta}=0$ and 
$\Gamma^m\tilde{D}_m\chi=0$, along with the gauge conditions 
$\partial^{\alpha}\psi_{\alpha}=0$ and $\gamma^{\alpha}\psi_{\alpha}=0$.  
Then, Eq. (\ref{grtneq}) reduces to the following equation satisfied by $u(y)$:
\begin{equation}
\Gamma^y\left[\partial_y-{{p+2n-1}\over{4(n-1)}}{{\cal W}^{\prime}\over
{\cal W}}\right]u=0,
\label{grtnkkzreq}
\end{equation}
from which we have $u\sim{\cal W}^{{p+2n-1}\over{4(n-1)}}$.  The bulk action 
(\ref{grtnact}) with the KK zero mode substituted takes the form
\begin{equation}
\int d^Dx\sqrt{-G}\,i\bar{\Psi}_M\Gamma^{MNP}D_N\Psi_p\sim
\int^{1/K_p}_{-1/K_p}dy{\cal W}^{-{1\over 2}}\int d^{p+1}x\,
i\bar{\psi}_{\alpha}\gamma^{\alpha\beta\delta}\partial_{\beta}\psi_{\delta}.
\label{effgrvtact}
\end{equation}
The integrand of the $y$-integral has the same form as that in the spinor 
field effective action (\ref{frmkkzrrm}).  So, the normalization and 
localization properties of the bulk gravitino KK zero mode are the same as 
those of the bulk spin 1/2 field discussed in the above.  Namely, the KK 
zero mode of the bulk spin 3/2 field is normalizable for any $a_p$ and 
the delocalized dilatonic $p$-brane with codimension greater than two 
($n>1$) localizes the bulk spin 3/2 field in a strict sense for any $a_p$.

\end{sloppypar}

\end{document}